\RequirePackage{fix-cm}
\documentclass[twocolumn,epjc3]{svjour3}  
\smartqed  
\RequirePackage{graphicx}
\RequirePackage{verbatim}
\usepackage{lineno}
\begin{document}
\title{Borexino's search for low-energy neutrinos associated with gravitational wave events from GWTC-3 database.}




 \author{
 D.~Basilico\thanksref{Milano}
 \and
 G.~Bellini\thanksref{Milano}
 \and
 J.~Benziger\thanksref{PrincetonChemEng}
 \and
 R.~Biondi\thanksref{MPIH,LNGS}
 \and
 B.~Caccianiga\thanksref{Milano}
 \and
 F.~Calaprice\thanksref{Princeton}
 \and
 A.~Caminata\thanksref{Genova}
 \and
 A.~Chepurnov\thanksref{Lomonosov}
 \and
 D.~D'Angelo\thanksref{Milano}
 \and
 A.~Derbin\thanksref{Peters,Kurchatov}
 \and
 A.~Di Giacinto\thanksref{LNGS}
 \and
 V.~Di Marcello\thanksref{LNGS}
 \and
 X.F.~Ding\thanksref{IHEP,Princeton}
 \and
 A.~Di Ludovico\thanksref{LNGSG,Princeton} 
 \and
 L.~Di Noto\thanksref{Genova}
 \and
 I.~Drachnev\thanksref{Peters}
 \and
 D.~Franco\thanksref{APC}
 \and
 C.~Galbiati\thanksref{Princeton,GSSI}
 \and
 C.~Ghiano\thanksref{LNGS}
 \and
 M.~Giammarchi\thanksref{Milano}
 \and
 A.~Goretti\thanksref{LNGSG,Princeton} 
 \and
 M.~Gromov\thanksref{Lomonosov,Dubna}
 \and
 D.~Guffanti\thanksref{Bicocca,Mainz}
 \and
 Aldo~Ianni\thanksref{LNGS}
 \and
 Andrea~Ianni\thanksref{Princeton}
 \and
 A.~Jany\thanksref{Krakow}
 \and
 V.~Kobychev\thanksref{Kiev}
 \and
 G.~Korga\thanksref{London,Atomki}
 \and
 S.~Kumaran\thanksref{CALI,Juelich,RWTH}
 \and
 M.~Laubenstein\thanksref{LNGS}
 \and
 E.~Litvinovich\thanksref{Kurchatov,Kurchatovb}
 \and
 P.~Lombardi\thanksref{Milano}
 \and
 I.~Lomskaya\thanksref{Peters}
 \and
 L.~Ludhova\thanksref{Juelich,RWTH}
 \and
 I.~Machulin\thanksref{Kurchatov,Kurchatovb}
 \and
 J.~Martyn\thanksref{Mainz}
 \and
 E.~Meroni\thanksref{Milano}
 \and
 L.~Miramonti\thanksref{Milano}
 \and
 M.~Misiaszek\thanksref{Krakow}
 \and
 V.~Muratova\thanksref{Peters}
 \and
 R.~Nugmanov\thanksref{Kurchatov,Kurchatovb}
 \and
 L.~Oberauer\thanksref{Munchen}
 \and
 V.~Orekhov\thanksref{Mainz}
 \and
 F.~Ortica\thanksref{Perugia}
 \and
 M.~Pallavicini\thanksref{Genova}
 \and
 L.~Pelicci\thanksref{Juelich,RWTH}
 \and
 \"O.~Penek\thanksref{GSI,Juelich}
 \and
 L.~Pietrofaccia\thanksref{LNGSG,Princeton}
 \and
 N.~Pilipenko\thanksref{Peters}
 \and
 A.~Pocar\thanksref{UMass}
 \and
 G.~Raikov\thanksref{Kurchatov}
 \and
 M.T.~Ranalli\thanksref{LNGS}
 \and
 G.~Ranucci\thanksref{Milano}
 \and
 A.~Razeto\thanksref{LNGS}
 \and
 A.~Re\thanksref{Milano}
 \and
  N.~Rossi\thanksref{LNGS}
 \and
 S.~Sch\"onert\thanksref{Munchen}
 \and
 D.~Semenov\thanksref{Peters}
 \and
 G.~Settanta\thanksref{ISPRA,Juelich}
 \and
 M.~Skorokhvatov\thanksref{Kurchatov,Kurchatovb}
 \and
 A.~Singhal\thanksref{Juelich,RWTH}
 \and
 O.~Smirnov\thanksref{Dubna}
 \and
 A.~Sotnikov\thanksref{Dubna}
 \and
 R.~Tartaglia\thanksref{LNGS}
 \and
 G.~Testera\thanksref{Genova}
 \and
 E.~Unzhakov\thanksref{Peters}
 \and
 A.~Vishneva\thanksref{Dubna}
 \and
 R.B.~Vogelaar\thanksref{Virginia}
 \and
 F.~von~Feilitzsch\thanksref{Munchen}
 \and
 M.~Wojcik\thanksref{Krakow}
 \and
 M.~Wurm\thanksref{Mainz}
 \and
 S.~Zavatarelli\thanksref{Genova}
 \and
 K.~Zuber\thanksref{Dresda}
 \and
 G.~Zuzel\thanksref{Krakow}
 }
 \thankstext{MPIH}{Present address: Max-Planck-Institut f\"ur Kernphysik, 69117 Heidelberg, Germany}
\thankstext{IHEP}{Present address: IHEP Institute of High Energy Physics, 100049 Beijing, China}
\thankstext{LNGSG}{Present address: INFN Laboratori Nazionali del Gran Sasso, 67010 Assergi (AQ), Italy}
\thankstext{Bicocca}{Present address: Dipartimento di Fisica, Universita degli Studi e INFN Milano-Bicocca, 20126 Milano, Italy}
\thankstext{CALI}{Present address: Department of Physics and Astronomy, University of California, Irvine, California, USA}
\thankstext{GSI}{Present address: GSI Helmholtzzentrum f\"r Schwerionenforschung GmbH, 64291 Darmstadt, Germany}
\thankstext{ISPRA}{Present address: Istituto Superiore per la Protezione e la Ricerca Ambientale, 00144 Roma, Italy}

 \institute{Dipartimento di Fisica, Universit\`a degli Studi e INFN, 20133 Milano, Italy\label{Milano}
 \and
 Chemical Engineering Department, Princeton University, Princeton, NJ 08544, USA\label{PrincetonChemEng}
 \and
  INFN Laboratori Nazionali del Gran Sasso, 67010 Assergi (AQ), Italy\label{LNGS}
 \and
  Physics Department, Princeton University, Princeton, NJ 08544, USA\label{Princeton}
 \and
 Dipartimento di Fisica, Universit\`a degli Studi e INFN, 16146 Genova, Italy\label{Genova}
 \and
  Lomonosov Moscow State University Skobeltsyn Institute of Nuclear Physics, 119234 Moscow, Russia\label{Lomonosov}
 \and
   St. Petersburg Nuclear Physics Institute NRC Kurchatov Institute, 188350 Gatchina, Russia\label{Peters}
 \and
 National Research Centre Kurchatov Institute, 123182 Moscow, Russia\label{Kurchatov}
 \and
  APC, Universit\'e de Paris, CNRS, Astroparticule et Cosmologie, Paris F-75013, France\label{APC}
 \and
  Gran Sasso Science Institute, 67100 L'Aquila, Italy\label{GSSI}
 \and
  Joint Institute for Nuclear Research, 141980 Dubna, Russia\label{Dubna}
 \and
 Institute of Physics and Excellence Cluster PRISMA+, Johannes Gutenberg-Universit\"at Mainz, 55099 Mainz, Germany\label{Mainz}
 \and
   M.~Smoluchowski Institute of Physics, Jagiellonian University, 30348 Krakow, Poland\label{Krakow}
 \and
 Institute for Nuclear Research of NAS Ukraine, 03028 Kyiv, Ukraine\label{Kiev}
 \and
  Department of Physics, Royal Holloway, University of London, Egham, Surrey, TW20 OEX, UK\label{London}
 \and
 Institute of Nuclear Research (Atomki), Debrecen, Hungary\label{Atomki}
 \and
 Institut f\"ur Kernphysik, Forschungszentrum J\"ulich, 52425 J\"ulich, Germany\label{Juelich}
 \and
    RWTH Aachen University, 52062 Aachen, Germany\label{RWTH}
 \and
  National Research Nuclear University MEPhI (Moscow Engineering Physics Institute), 115409 Moscow, Russia\label{Kurchatovb}
 \and
 Physik-Department, Technische Universit\"at  M\"unchen, 85748 Garching, Germany\label{Munchen}
 \and
  Dipartimento di Chimica, Biologia e Biotecnologie, Universit\`a degli Studi e INFN, 06123 Perugia, Italy\label{Perugia}
 \and
 Amherst Center for Fundamental Interactions and Physics Department, UMass, Amherst, MA 01003, USA\label{UMass}
  \and
 Physics Department, Virginia Polytechnic Institute and State University, Blacksburg, VA 24061, USA\label{Virginia}
 \and
 Department of Physics, Technische Universit\"at Dresden, 01062 Dresden, Germany\label{Dresda}
  }
\date{Received: date / Accepted: date}
\maketitle

\begin{abstract}
The search for neutrino events in correlation with gravitational wave (GW) events for three observing runs (O1, O2 and O3) from 09/2015 to 03/2020 has been performed using the Borexino data-set of the same period.
We have searched for signals of neutrino-electron scattering and inverse beta-decay (IBD) within a time window of $\pm1000$~s centered at the detection moment of a particular GW~event.
The search was done with three visible energy thresholds of $0.25$, $0.8$ and $3.0$~ MeV.
Two types of incoming neutrino spectra were considered: the mono-energetic line and the supernova-like spectrum.
GW candidates originated by merging binaries of black holes (BHBH), neutron stars (NSNS) and neutron star and black hole (NSBH) were analyzed separately.
Additionally, the subset of most intensive BHBH mergers at closer distances and with larger radiative mass than the rest was considered. 
In total, follow-ups of $74$ out of $93$ gravitational waves reported in the GWTC-3 catalog were analyzed and no statistically significant excess over the background was observed.
As a result, the strongest upper limits on GW-associated neutrino and antineutrino fluences for all flavors ($\nu_e, \nu_\mu, \nu_\tau$) at the level $10^9-10^{15}~\rm{cm^{-2}GW^{-1}}$ have been obtained in the $0.5 - 5$~MeV neutrino energy range.

\end{abstract}

\section{Introduction}\label{intro}

The era of multi-messenger astronomy has started with the detection of gravitational waves (GW) by the LIGO experiment~\cite{Abb2016}.
During O1 and O2 observing periods (09/2015 – 08/2017), LIGO/Virgo has detected $10$~binary black-hole mergers and a single binary neutron-star merger~\cite{Abb2019}.
Firstly, the short gamma-ray burst GRB170817A was detected in $1.7$~s temporal interval coincidence with the GW170817 event from the binary neutron-star merger~\cite{Abb2017}.
At present the third LIGO, Virgo and KAGRA Collaboration Gravitational Wave Transient Catalog (GWTC-3) consists of transient GW~signal records discovered up to the end of LIGO-Virgo's third observing run (03)~\cite{Abb2021}.

The observation of GW~events triggered an intensive follow-up campaign in neutrino detectors~\cite{Adr2016,Aab2016,Gan2016,Abe2016,Ago2017,Alb2017,Abe2018,Ace2020,An2021,Ace2021,Abe2021,Abe2021A,Abe2021B,Bol2021,Pet2021}.
\v Cerenkov neutrino telescopes (ANTARES, IceCube~\cite{Adr2016,Alb2017}) and Pierre Auger Observatory~\cite{Aab2016} have searched for high energy neutrinos above $100$~GeV and $100$~PeV, respectively.
The experiment KamLAND has searched for inverse beta decay (IBD) antineutrino events within $(1.8 - 111)$~MeV energy range~\cite{Gan2016,Abe2021A} and the Super-Kamiokande collaboration has reported results for neutrino signals within $3.5$~MeV to $100$~PeV~energy range~\cite{Abe2016,Abe2018,Abe2021B}.
The Borexino detector has searched for correlated neutrino events with visible energies above $0.25$~MeV within a $\pm 500$~s time window centered at the detection moment of the first three reported GW~events in assumption of monochromatic and Fermi-Dirac spectra~\cite{Ago2017}.
The Daya~Bay collaboration has searched for possible electron-antineutrino signals with energies from $1.8$ to $100$~MeV in coincidence with $7$~GW~events including GW~170817 within three time windows of $\pm 10$, $\pm 500$, and $\pm 1000$~s relative to the occurrence of the GW~events~\cite{An2021}. 
The XMASS-I $832$~kg xenon detector has searched for event bursts associated with the $11$~GW~events detected during LIGO/Virgo’s O1 and O2 periods~\cite{Abe2021}.
The NOvA neutrino detectors have performed search for any signal coincident with $28$~GW~events and supernova like neutrino interactions in coincidence with $76$~GW~events~\cite{Ace2020,Ace2021}.
The Baksan Underground Scintillation Telescope has searched for $\nu_\mu$ and $\bar{\nu}_\mu$ with energies above $1$~GeV from the directions of the localization of the GW~events and in temporal coincidence with the GW170817 occurring due to the merger of two neutron stars~\cite{Pet2021,Bol2021}.

The neutrino and antineutrino events within a time window of $\pm 1000 (\pm 500, \pm 10)$ seconds around the moment of gravitational wave detection were analyzed in the detectors mentioned above, but no evidence for an excess of coincident neutrino events had been reported.
Combination of the data from gravitational, neutrino and electromagnetic detectors forms a new multi-messenger approach leading to a deeper understanding of astrophysical and cosmological processes through combination of information from different probes.

Here, we report the results of a search for signals with visible energy above $0.25$~MeV in the Borexino detector in coincidence with GW~events from GWTC-3.
We look for neutrino signals from $\nu_e, \nu_{x=\mu, \tau}$ and antineutrinos $\bar{\nu}_e, \bar{\nu}_{x=\mu,\tau}$ originated in the GW~events that scatter on electrons.
We also search for signals of $\bar{\nu}_e$ that induce IBD reaction on protons.

Two different spectra of incoming neutrinos ($\nu_{e,x}$ and $\bar{{\nu}}_{e,x}$) were used for the analysis: the mono-energetic line and the spectrum expected from supernovae.
The same $\bar{{\nu}}_{e}$-neutrino spectra were considered for search with the IBD reaction.
The temporal correlation analysis between Borexino events and GW~events were performed for three different merging modes of black holes and neutron stars - BHBH, NSNS, NSBH. 

Negative results of searches for neutrino radiation accompanying GW~events are presented as a limit on the neutrino or antineutrino fluences.
In the papers above, the results of various detectors and different numbers of GW~events were analyzed, sometimes the obtained limits of the neutrino fluence are normalized to the number of GW~events.
Since the data on the radiation mass and the distance to the event $R_i$ are available from the GWTC-3 catalog for almost all events, it would be natural, in assumption that the neutrino fluence has to be proportional to the radiation mass and inversely proportional to the square of $R_i$, to use this factor for comparison of the different experiments results.



\section{The Borexino Detector}
The Borexino is a liquid scintillator-based large volume detector specifically designed for neutrino detection. The experiment was located at the Laboratori Nazionali del Gran Sasso (LNGS) at the depth of $3800$~meters of water equivalent and has been operated since from May 2007 till October 2021. Such location provided good cosmic muon flux suppression by a factor of $\approx 10^6$. The detector structure represents an implementation of the graded shielding concept. 

The water tank (WT) is constructed of stainless steel with high radiopurity and contains $2100$~tons of ultra-pure water as additional shielding imposed to suppress external $\gamma$-rays and neutrons.
The WT contains a stainless steel sphere (SSS) with radius of $6.75$~m and thickness of $8$~mm that serves as the scintillating inner detector body.
The WT is equipped with~$208$~8-inch PMTs placed on its floor and outer surface of the SSS and is used as the Cherenkov muon veto (outer detector, OD) for identification of residual muons crossing the detector.
The scintillation light is detected by nominally~$2212$~8-inch PMTs of~the inner detector (ID) uniformly distributed on~the inner surface of~the~SSS.
The neutrino target consists of $278$~tons ultra-pure organic liquid scintillator and is confined in the innermost detector part, that is divided by two 125 $\mu$m nylon "balloons", the Radon Barrier (RB) and the Inner Vessel (IV) with radii of $5.50$~m and $4.25$~m respectively.

The scintillator was chosen for the purpose of low-energy neutrino registration as pseudocumene (PC 1,3,4-trimethylbenzene, $\rm{C_6H_3(CH_3)_3}$) doped with a fluorescent constituent PPO (2,5-diphenyloxazole, $\rm{C_{15}H_{11}NO}$) in concentration of $1.5$~g/l.
The buffer volume serves as neutron and gamma radiation shield and is posed with a solution of a light quencher dope consisting of $2$~g/l dimethylphthalate (DMP, $\rm{C_6H_4(COOCH_3)_2}$). 

Detection of charged particles in the Borexino detector occurs via production of the scintillator light in the scintillator volume and its detection by the PMTs. 
Since the moment the detector started operation until the end of the experiment, the number of active PMTs has decreased that is taken into account in the current analysis. 
The data are used for reconstruction of energy and spatial coordinates of an event and also allows identification a particle type $(e,\alpha,\mu)$ due to differences in the scintillation time profiles. 

Both energy and spatial resolutions of the detector were studied with radioactive sources placed at different positions inside the inner vessel. 
The energy and position resolutions are $\sigma_E \approx 50$~keV and $\sigma_X \approx 10$~cm at $1$~MeV with $2000$~PMTs, respectively; both are scaling with the energy of an event as $\sim 1/\sqrt{E}$ at low energies.
The primary electronics of the Borexino detector is optimized for energies up to few MeV with energy calibration reliable up to $16.8$~MeV. 
For the purpose higher energies operation the system of $96$ fast waveform $400$~MHz digitizers was developed, each of them is reading-in the signal summed from $24$~PMTs.
The Borexino detector is unable to determine directional information of a single event due to nearly isotropic emission of scintillation light (see, however,~\cite{Ago2022A,Ago2022B}).

A more detailed description of the Borexino detector can be found in the following papers~\cite{Ali2002,Ali2009,Bac2012,Bel2014,Ago2019}.

Neutrinos $\nu_{e,x}$ and antineutrinos and $\bar{{\nu}}_{e,x}$ are detected by means of their elastic scattering on electrons:
\begin{equation}\label{eq:1}
    \nu_{e,x}~(\bar{{\nu}}_{e,x})~+~e^- \rightarrow \nu_{e,x}~(\bar{{\nu}}_{e,x})~+~e^-.
\end{equation}
For a given (anti)neutrino energy $E_\nu$, the maximum electron recoil energy $ E_{emax}$ is given by the formula $E_{emax}=2E_\nu^2/(2E_\nu+m_e)$
where $m_e$ is the electron mass.

Electron antineutrinos $\bar{\nu}_e$ can also be detected via the inverse beta-decay (IBD) reaction with an energy threshold of $1.8$~MeV:
\begin{equation}\label{eq:3}
    \bar{\nu}_e+p \rightarrow n + e^+.
\end{equation}
The visible energy of the positron and two annihilation photons is related to the antineutrino energy as $E_{vis} = E_{\bar{\nu}_e} -~0.784~$MeV.
The neutron capture on protons produces a delayed $2.22$~MeV $\gamma$'s with mean capture time of $\sim 260$~$\mu$s~\cite{Bel2011d}.

The Borexino was the first experiment that has detected and then precisely measured all solar neutrino fluxes (except for the $hep$-neutrino)~\cite{Arp2008,Bel2011a,Bel2010,Ago2020,Bel2012,Bel2014a,Ago2018} and has also registered antineutrinos emitted in decay of radionuclides naturally occurring within the Earth~\cite{Bel2010g,Bel2013g,Ago2015,Ago2020b}.
The Borexino detector is perfectly suited for the study of other fundamental problems, as well as searching for rare and exotic processes in particle physics and astrophysics~\cite{Bel2012a,Bel2011c,Ago2021,Aga2020,Bel2010a,Bel2012b,Bel2013,Ago2015a}.
Additionally, temporal correlations with transient astrophysical sources such as $\gamma$-ray bursts~\cite{Ago2017a}, gravitational wave events~\cite{Ago2017}, solar flares~\cite{Ago2021}, and fast radio burst~\cite{App2022} have been performed.


\section{The Borexino Data Selection}
\label{Bx Data}
The aim of data selection is to provide maximum exposure for the desired study with minimum background contribution. 
In the current analysis we search for neutrino-electron scattering, a reaction that has no specific interaction signature.
Thus, the background has to be suppressed generically, as a reduction of the detector count rate per unit of exposure.
Background composition of the Borexino experiment was carefully studied in the course of many years of research. 
In the current study we account for the following background component groups:
\begin{itemize}
	\item{Short-lived cosmogenic backgrounds ($\tau \leq 1$~s) such as $^{12}\rm{B}$, $^{8}$He, $^{9}$C, $^{9}$Li etc., and long-lived cosmogenic backgrounds ($\tau \geq 1$~s) such as $^{11}$Be, $^{10}$C, $^{11}$C etc., produced within the detector fiducial volume.}
	\item{External backgrounds present in the material of the inner nylon vessel such as $^{210}$Pb and Uranium/Thorium decay chains.}
	\item{Internal natural backgrounds contained in the bulk of the detector fluid such as $^{14}$C, $^{85}$Kr, $^{210}$Bi and $^{210}$Pb.}
\end{itemize}  

These backgrounds can be suppressed by using information coming from the processed detector data such as spatial event distributions or ID/OD temporal and spatial coincidences.
Cosmogenic backgrounds are reduced by applying the detector temporal veto after each muon, which could be discriminated through coincidence with outer veto as well as by pulse-shape discrimination~\cite{Bel2011d}.
A veto duration of $0.3$~s after each muon crossing the IV is applied to suppress $^{12}$B to a statistically insignificant level and reduce $^{8}$He, $^{9}$C and $^{9}$Li by factor of $3$ with a live time loss as small as $1$~\%. 
More long-lived cosmogenic backgrounds of $^{11}\rm{C}$, $^{10}\rm{C}$, $^{11}\rm{Be}$ and others can were not specifically suppressed by advanced veto system in order to save maximal exposure for neutrino events.

Backgrounds contained in the bulk of the detector can not be discriminated on event basis since they can not be localized neither spatially nor temporally.
Nevertheless, the count rate of these background components are reduced by setting a cut on visible energy.
This is important specifically due to the presence of $^{14}\rm{C}$ in the scintillator.
$^{14}\rm{C}$ produces a beta-spectrum with an endpoint of $156$~keV and has activity of roughly $110$~Bq in the whole inner vessel.

The presence of this spectral component sets the lower threshold of the analysis to $E_1 = 0.25$~MeV of visible energy\footnote{Visible energy spectrum of $^{14}$C is broadened up to this value due to energy resolution of the detector}.
Two additional thresholds of $E_2 = 0.8$ and $E_3 = 3$~MeV of visible energy are applied for higher energy neutrino search. 
The second $0.8$~MeV threshold was set in order to exclude $^{210}\rm{Po}$ and solar $^7$Be neutrino events and the third $3$~MeV threshold rejects the most part of natural radioactivity.
The threshold of $0.25$~MeV allows to register neutrinos with energy as low as $0.4$~MeV via $(\nu,e)$-elastic scattering.

Backgrounds contained in the nylon of IV can not be removed by any kind of purification and are therefore of the order of $10^2$ -- $10^3$ times higher than those within the bulk of the scintillator.
The most important contributions come from $^{214}$Bi and $^{208}$Tl decays.
These nuclides undergo $\beta$ and $\beta+\gamma$ decay processes with a continuous spectrum overlapping with the region of interest of the current analysis.
The most effective way to suppress this kind of backgrounds is to perform a geometrical cut on events, selecting those within a fiducial volume. 
The fiducial volume is defined in such manner that all events within and further than $75$~cm away from the IV are kept.
The distance of $75$~cm corresponds to $3$ standard deviations of position reconstruction uncertainty at the lowest $0.25$~MeV energy threshold\footnote{Position reconstruction precision increases with energy due to statistical reasons}. 
The corresponding fiducial volume has a mass of $145$~t.


\section{GW~events, temporal window and neutrino spectra}

We have used the GWTC-3 database compiled by the LIGO and VIRGO collaborations for O1, O2 and O3 observing runs~\cite{Abb2021}. 
This database contains information about possible sources of GW~events, merger time of the event in GPS seconds, mass $M1$ and $M2$, chirp mass $\delta M$, final mass $M_f$ in solar mass units, red-shift $z$, and distance information $R$.

During the period of interest from September 2015 to March 2020, $93$~GW~events have been observed, $87$ of which are black hole mergers(BHBH), 2 events of neutron star merge (NSNS) and 4 events of neutron star - black hole merge (NSBH).

The nearest GW~event is the most famous merger of two neutron stars GW170817 occurred at $R=40^{+7}_{-15}$~Mpc.
The biggest redshift $z = 1.18^{+0.73}_{-0.53}$ ($R=8280^{+6720}_{-4290}$~Mpc) was observed for GW190403-051519. 
This was the only GW~event with red shift $z > 1$, while 90\% of the GW~events have $z  < 0.64$.

We have considered the coincidence time window $\Delta t = 2000$~s centered at the GW observation time with a width of $\pm1000~$s covering a possible delay of sub-MeV neutrinos propagating at the sublight speed.
Negative interval of time window $\Delta t = 0..-1000$~s covers earlier emission of neutrinos in the case of binary mergers \cite{Bar2011}.
For a distance corresponding to the $z = 0.64$ redshift, the delay will reach $1000$~s in case of $0.6$~MeV neutrinos with a rest mass of $65$~meV, which is the upper limit on the heaviest neutrino mass state from the Planck 2018 data~\cite{Agh2018} and oscillation mass squared differences~\cite{Est2019}. 
All selected GW~events had the data taking time above $95$\% of the corresponding time interval $\Delta t$.

There's no expectation of MeV neutrino emission from BHBH mergers, since no ordinary matter is present. 
Neutrinos could only be produced either by some exotic scenario or if the GW event were misidentified and was actually either a supernova or a merger including neutron star. 
Since there is no reliable theory for the low-energy part of the neutrino emission spectrum for BHBH mergers, we calculated fluence limits for two different kinds of possible neutrino spectra: the monoenergetic line and the supernovae low-energy continuous spectrum $\Phi(E_\nu)$.
The latter was assumed to be quasi-thermal spectrum with mean energy $\langle E \rangle$ and deviation from thermal distribution characterized by the pinching parameter $\alpha = 3$ for all neutrino flavors ($\nu_{e,x}, \bar{\nu}_{e,x}$)~\cite{Tam2012,Luj2014,Mir2016}.
The emitted neutrino spectrum $S(E_\nu)$ depends on the neutrino energy $E_\nu$ as:
\begin{equation}\label{eq:5}
\Phi(E_\nu) \sim (E_\nu/T)^\alpha e^{(-E_\nu/T)},
\end{equation} 
where $T = \langle E \rangle / (\alpha + 1)$ is effective temperature, which was considered to be the same for all neutrino flavors.

The expected number of events depends on neutrino spectrum and $(\nu,e)$-scattering cross section $d\sigma(E_\nu)/dE_e$~\cite{Bah1995}.
The total cross section $\sigma(E_a,E_b)$ for electron with energy in the range $(E_a,E_b)$  is obtained by integrating the $d\sigma(E_\nu)/dE_e$ over recoil electron energies $E_e$ between the electron energies $E_a$ and $E_b$:
\begin{equation}
\sigma(E_a,E_b) = \int\frac{d\sigma(E_\nu,E_e)}{dE_e}dE_e
\label{CSmono} 
\end{equation}
If the neutrino spectrum $\Phi(E_\nu$) is not a monochromatic line, the total cross section for the electron recoil energy interval $(E_1,E_2)$ is calculated as:
\begin{equation}
\sigma(E_a,E_b) = \int\int\frac{d\sigma(E_\nu,E_e)}{dE_e}\Phi(E_\nu)dE_edE_\nu
\label{CSspectr}
\end{equation}
Additionally, in order to compare the theoretical cross sections (\ref{CSmono}) and (\ref{CSspectr}) with the experimental results, finite energy resolution of the detector has to be taken into account~\cite{Ago2017a}.
The neutrino–electron cross sections, the number of electrons in the Borexino fiducial volume and the GW neutrino fluences will determine the expected number of neutrino events in the detector.

\section{Analysis of the temporal correlations of Borexino signals with GW~events}

The Borexino detector was in data taking mode when 70 (out of 87) black hole mergers, 2 neutron star mergers and 2 (out of 4) neutron star black hole mergers  have been occurred.
Three different variants of binary merges in temporal coincidence with the Borexino data were analyzed separately.
Some parameters of GW~events involving neutron stars as obvious possible source of the neutrino flux are shown in Table~\ref{tab:1}.
Additionally, average values of the these parameters for 70 BHBH mergers are shown in the bottom row of Table~\ref{tab:1}.

\begin{table}[ht]
\caption{GW~events involving neutron stars, for which the temporal correlation analysis was carried out with the Borexino signals. $M1$ and $M2$ are masses in solar mass units $M_\odot$, $R$ is distance in Mpc, $M^{rad}$ is radiative mass in $M_\odot$ units. Average values of $M1$, $M2$, $R$ and $M^{rad}$ are given for 70 black hole mergers. }
\label{tab:1}
\begin{tabular}{|c|c|c|c|c|c|}
\hline
GW~event & Mode & $M1$  & $M2$ & $R$ & $M^{rad}$\\
\hline
GW 170817 & NSNS & 1.46  & 1.27 & 40 & $\leq 0.04$ \\
GW 190425 & NSNS & 2.0 & 1.40 & 160 & -- \\ 
GW 190426 & NSBH & 5.7 & 1.5 & 370 & -- \\ 
GW 191219 & NSBH & 31.1 & 1.17 & 550 & 0.1 \\ 
70 GW BHs  & BHBH & 36.7 & 23.0 & 2130 & 2.4 \\ 
\hline
\end{tabular}
\end{table}

For the possible registration of neutrinos, the most interesting is the GW170817 signal measured on 2017 August 17 produced by the coalescence of two neutron stars with masses $1.46~M_\odot$ and $1.27~M_\odot$  and occurred at a record close distance $R=40^{+7}_{-15}$~Mpc. 
The second registered neutron star merger GW190425 event occurred at a distance 4 times greater with the close chirp mass, which determines the amplitude of the GW signal.

\begin{figure}[ht]
    \centering
 \includegraphics[width=.9\linewidth]{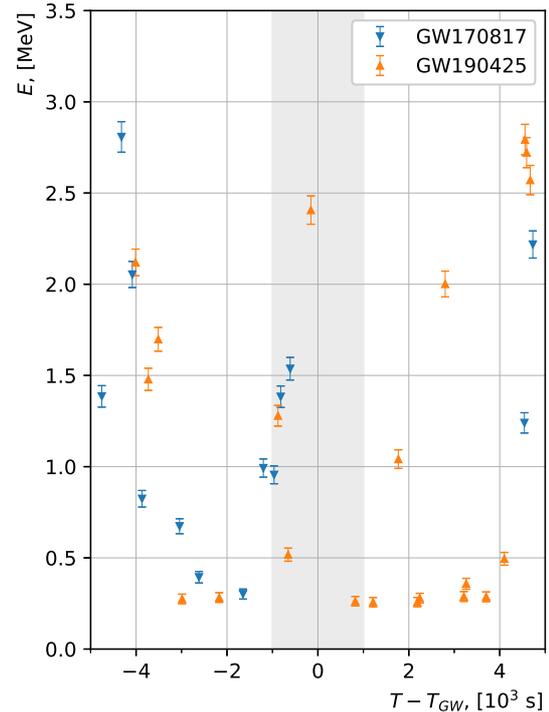}
	\caption{Borexino events with energies above $0.25$~MeV occurring within $\pm5000$~s of GW170817 (blue inverted triangles) and GW190425 (yellow triangles) events produced by the coalescence of two neutron stars.
	All events are consistent with the expected rates of solar neutrinos and background events.}
	\label{fig:1}
\end{figure}

\begin{figure}[ht]
    \centering
    \includegraphics[width=.9\linewidth]{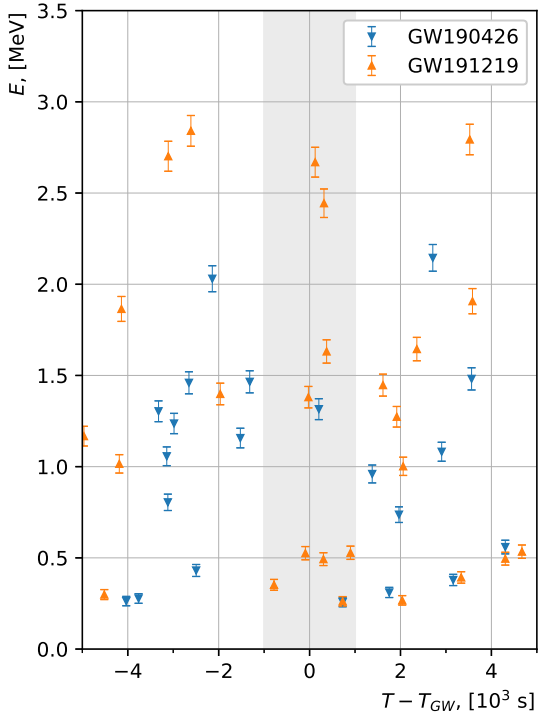}
	\caption{Borexino events with energies above $0.25$~MeV occurring within $\pm 5000$~s of GW190426 (blue inverted triangles) and GW191219 (yellow triangles) events produced by the coalescence of neutron star and black hole.
	}
	\label{fig:2}
\end{figure}

The detection time and energy of  Borexino events passing all data selection cuts in $\pm~5000$~s windows around GW170817 and GW190425 events due to the merger of two neutron stars are shown in Figure~\ref{fig:1}.
The closest events with energies 1.54 (2.41) MeV occurred at 610 (154) s before GW170817 (GW190425), respectively.
A similar Figure~\ref{fig:2} shows  Borexino events for two GW190426 and GW191219 events corresponding to the merger of a neutron star and a black hole.

Number of  Borexino events in $\pm~1000$ s interval with energy above $0.25$~MeV ($N_1$) and $0.8$~MeV ($N_2$) in comparison with the reduced backgrounds defined for intervals $[-5000 \ldots -1000, 1000 \ldots 5000]$~s are shown in Table~\ref{tab:2} for 4~GW~events involving neutron stars.
One can see that no excess of the counting rate, associated with GW~events involving neutron stars, above the expected background is observed.
So, there are only $3$~events in the $\pm1000$~s interval  centered at the GW170817 event arrival time with the energy in the $0.25 - 16.8$~MeV range, while $2.5 \pm 0.8$ solar neutrino and background reduced events were obtained within the $[-5000 \ldots -1000,1000 \ldots 5000]$~s window. 
All detected events were in agreement with expected solar neutrino and background count rates.

\begin{figure}[ht]
\includegraphics[width=.9\linewidth]{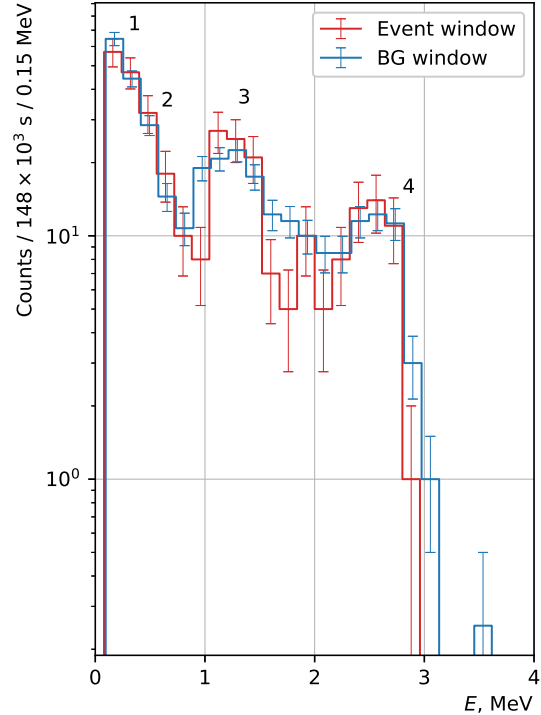}
	\caption{Borexino visible energy spectrum of singles in correlation with 70~BHBH GW~events in the $\pm1000$~s time window (red line with errors). Blue line  shows the normalized background spectrum measured in$\left[{-5000} \dots {-1000}\right]$~s and $\left[{1000} \ldots {5000}\right]$~s intervals. 1 - $^{210}\rm{Po}~~ \alpha$-peak, 2 – recoiled electrons from the solar $^7\rm{Be}$-neutrino, 3 – $^{11}\rm{C}~~ \beta^+$-decay, 4 – external background ($^{208}\rm{Tl}$).}
\label{fig:3}
\end{figure}

The energy spectrum of  Borexino events in correlation with BHBH GW~events in the $\pm1000$~s time window for $150$~keV interval and the normalized background spectrum of events registered in $\left[{-5000} \dots {-1000}\right]$~s and $\left[{1000} \ldots {5000}\right]$~s intervals are shown in Figure~\ref{fig:3}.

Number of events and the reduced background in $\pm~1000$~s interval with energy above $0.25$~MeV and $0.8$~MeV for $70$~GW BHBH events are shown in Table~\ref{tab:2}.
No statistically significant excess of the difference between these spectra for any energy interval was observed.

The spectrum in Figure~\ref{fig:3} is dominated by $^{14}$C in the region below $0.250$~MeV of visible energy, by the recoil electrons from solar $^7$Be neutrinos in $0.25 - 0.8$~MeV interval, by cosmogenic $^{11}$C in $1 - 2$~MeV region and by external gamma-quanta of $^{214}$Bi and $^{208}$Tl in $2 - 3$~MeV region.
All these components can not be significantly reduced by any available data selection techniques without serious exposure loss.
Figure~\ref{fig:3} shows the energy spectrum in the range 0.25-4.0 MeV and only three events were detected with energies above 4 MeV.


\begin{table}[ht]
\caption{Number of events $N_1$ and $N_2$ in $\pm~1000$ s interval with energies above $0.25$~MeV and $0.8$~MeV, correspondingly, for GW~events involving neutron stars. For comparison the reduced backgrounds $B_1$ and $B_2$ defined for intervals $[-5000..-1000,1000..5000]$~s are also shown. The same data for 70~BHBH GW~events is shown in the last column.}
\label{tab:2}
\begin{tabular}{|c|c|c|c|c|c|}
\hline
GW & 170817 & 190425  & 190426 & 191219 & 70 GWs\\
\hline
mode & NSNS & NSNS  & NSBH & NSBH & BHBH \\
\hline
$N_1$ & 3 & 4 & 2 & 9 & 304 \\ 
$B_1$ & 2.5$\pm$0.8 & 4.3$\pm$1.0 & 4.8$\pm$1.1 & 4.3$\pm$1.1 & 310$\pm$9 \\ 
$N_2$ & 3 & 2 & 1 & 4 & 158 \\ 
$B_2$ & 1.8$\pm$0.7 & 2.0$\pm$0.7 & 2.8$\pm$0.8 & 3.0$\pm$0.9 & 163$\pm$7\\ 
\hline
\end{tabular}
\end{table}

Based on the spectrum of the Borexino detector in Figure~\ref{fig:3} three mentioned above energy thresholds $E_{1,2,3}$ are selected for the analysis. 
The analysis is performed for the energy intervals from $E_{1,2,3}$ to $E_{emax}$, the latter must not exceed the range of validity of the detector energy response calibration (0.25 - 16.8) MeV.
In the analysis, the energy resolution of the detector $\sigma(E_e)$ is taken into account~\cite{Ago2017a}.
We calculated the overall number of candidate events above $E_{1,2,3}$ in the $\pm 1000$~interval for various values of neutrino energy $E_\nu$ or average energy $<E>$ for supernova spectra.

Since there is no statistically significant excess of the number of events within $\pm 1000$~s GW window above the background, the upper limits on fluences $\Phi_{\nu_{e,x},\bar\nu_{e,x}}$ for mono-energetic (anti)neutrinos with the energy $E_\nu$ are calculated as:
\begin{equation}\label{eq:6}
    \Phi_{\nu_{e,x},\bar\nu_{e,x}} =
        \frac {N_{90}(E_\nu, n_\mathrm{obs}, n_\mathrm{bkg})} {r N_e \sigma(E_\mathrm{th}, E_{e_\mathrm{max}})},
\end{equation}
where $E_{th} = E_{1,2,3}$, $N_{90} (E_\nu, n_{obs}, n_{bkg})$ is the 90\%~C.L. upper limit for the number of GW-correlated events in the $(E_{th}, E_{emax})$ interval per single GW and $N_e$ is the number of electrons in $145$~t of the Borexino scintillator.
The limits were obtained in assumption that the whole neutrino fluence consists of only one individual flavor.
The factor $\sigma(E_{th}, E_{emax})$ represents the cross section for detected neutrinos $(\nu_{e,x}, \bar\nu_{e,x})$ with the energy $E_{\nu}$ without oscillations while recoil electrons are detected in the interval ($E_{th}, E_{emax}$) taking into account the detector energy resolution~\cite{Ago2017a}. 
The recoil electron detection efficiency $r$ was taken as 1, with the accuracy corresponding to the precision of the fiducial volume definition ($\simeq1\%$ \cite{Ago2018}).

The numerator $N_{90}(E_\nu,n_{obs},n_{bkg})$ was calculated for spectra of coincident (in $\pm1000$ s) and non-coincident events in the energy interval ($E_{th}, E_{emax}$).
Here, $n_{obs}$ and $n_{bkg}$ denote overall numbers of observed and background events in this energy interval normalized by their respective times. 
The longer interval $\Delta t_{bkg}$ =18~ks for background detection was chosen in order to reduce the error of $n_{bkg}$ that plays an important role in the analysis.
The value $n_{bkg}$ was normalized by the overall time ratio taking into account the actual live time of the detector within these time windows. 

The procedure was repeated for neutrino energies $E_{\nu}$ from 0.5~MeV to 50~MeV in increments of 0.5 MeV. 
As was mentioned above, in order to have the best ratio of the expected effect with respect to the background and taking into account the shape of the spectrum (Figure~\ref{fig:3}) the three energy thresholds $E_1$, $E_2$ and $E_3$ were used for different neutrino energies. 
The analysis is performed in the energy interval  $(E_{1,2,3},~E_{emax})$ if $E_{emax}$ is not exceeded the range of the detector energy calibration, in the later case the value 16.8 MeV was used for the right border of the analysis interval.

\begin{figure}
    \includegraphics[width=9cm, height=10cm]{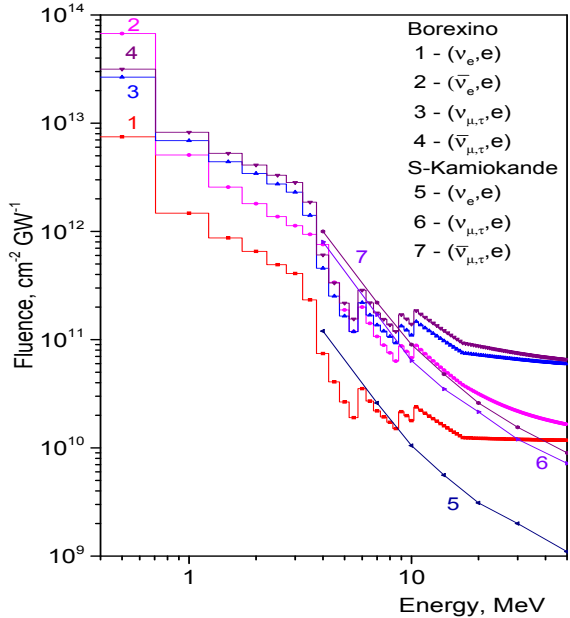}
	\caption{ Borexino upper limits on mono-energetic neutrino fluences obtained with $(\nu,e)$ elastic scattering reaction through the temporal correlation analysis for 74 GW~events (90\% C.L.): 1-$\nu_e$, 2-$\bar{\nu}_e$, 3-$\nu_{\mu,\tau}$,  4-$\bar{\nu}_{\mu,\tau}$. The limits 5, 6 and 7 of Super-Kamiokande obtained for single GW170817 event from $(\nu,e)$-scattering reaction are also shown~\cite{Abe2018}.}
	\label{fig:4}
\end{figure}

For the case of all $74$~GW~events the upper limits on neutrino and antineutrino fluences of different flavors normalized per single GW are shown in Figure ~\ref{fig:4}. 
The jumps in the upper limit at the energies above 7 MeV is associated with the inclusion of the above-mentioned three events with energies above 4 MeV in the analysis.
The limits change weakly at higher neutrino energies above 17 MeV due to the fact that the maximum energy of recoil electrons $E_{emax}$ is constrained by the value $16.8~\rm{MeV}$  in the analysis.
These are the first constraints on the fluence of neutrino with energy below 4~MeV obtained from the neutrino-electron scattering reaction. 
The previous limits obtained by Super-Kamiokande collaboration for the  GW170817 NSNS merger with $(\nu_{e,x},e)$- and $(\bar{\nu}_x,e)$-scattering are also shown in the Figure~\ref{fig:4}~\cite{Abe2018}.


For the case of supernovae spectrum with a mean energy with a mean energy $\langle E \rangle = 15.6$~MeV and the parameter $\alpha = 3$ (equation \ref{eq:5}) and integrating over the analyzed electron recoil energy interval $E_{1,2,3} - 16.8$~MeV, we get the limits on the total electron neutrino fluence per single GW: $\Phi(\nu_e) \leq 1.2 \times 10^{10} \rm{cm^{-2}}$ (90\%~C.L.) that is close to the limit obtained for mono-energetic neutrinos with the same energy.
The values of the limits on other neutrino flavors obtained from the $(\nu,e)$-scattering channel are given in Table~\ref{tab:3}. 
We also calculated the upper limit on the electron antineutrinos ($\bar{\nu}_e$) fluence given in Table~\ref{tab:3} using the IBD reaction (see section~\ref{section:IBD}).

\begin{figure}
\includegraphics[width=9cm, height=10cm]{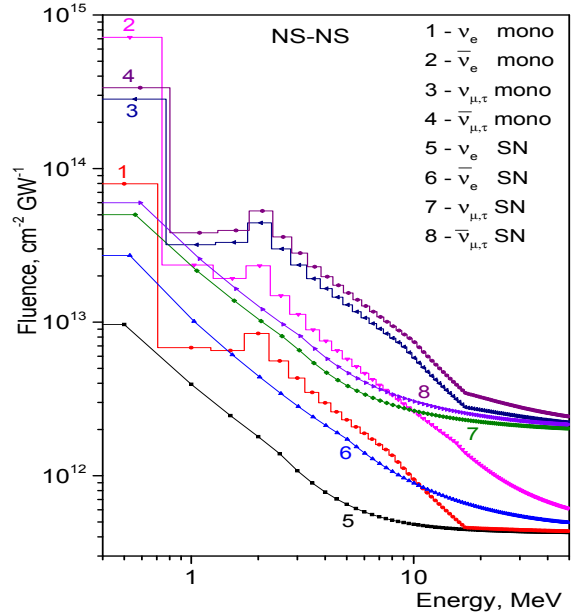}
	\caption{The merger of two neutron stars - GW170817 event. The upper limits on the fluences of monoenergetic neutrinos (mono): 1-$\nu_e$, 2-$\bar{\nu}_e$, 3-$\nu_{\mu,\tau}$,  4-$\bar{\nu}_{\mu,\tau}$ and neutrinos with supernova spectra (SN): 5-$\nu_e$, 6-$\bar{\nu}_e$, 7-$\nu_{\mu,\tau}$,  8-$\bar{\nu}_{\mu,\tau}$ (90\% C.L.). The x-scale shows the energy of monoenergetic neutrinos $E_\nu$ or the average energy $\langle E \rangle$ in the case of a supernova spectrum.}
	\label{fig:5}
\end{figure}

\begin{table}[ht]
\caption{Upper limits on fluences from all 74 events per single GW for all neutrino flavors obtained from the temporal correlation analysis in $10^{9}~\rm{cm^{-2}}$ units (90\%~C.L.) calculated for mono-energetic neutrinos and the supernova spectrum with $\langle E \rangle = 15.6$~MeV.
The limits in the right column obtained through the study of IBD reaction.}
\label{tab:3}
\begin{tabular}{|c|c|c|c|c|c|}
\hline
$E_{\nu}$ & ${\Phi}_{\nu_e}$ & ${\Phi}_{\bar{\nu}_{e}}$ & ${\Phi}_{\nu_{\mu,\tau}}$ & ${\Phi}_{\bar{\nu}_{\mu,\tau}}$ & ${\Phi}_{\bar{\nu}_{e}}$IBD\\
\hline
2 & $650$ & $1800$  & $ 3400$ & $4100$ & $100$ \\
6   & $35$ & $200$ & $220$ & $290$ & $1.7$ \\ 
10   & $18$ & $68$ & $110$ & $140$ & $0.51$ \\ 
14   & $16$ & $52$ & $97$ & $120$ & $0.25$ \\ 
18   & $12$ & $36$ & $74$ & $91$ & $-$ \\ 
30   & $12$ & $22$ & $66$ & $75$ & $-$ \\ 
50   & $12$ & $17$ & $60$ & $66$ & $-$ \\ 
$<15.6>$ & $12$ & $19$ & $63$ & $71$ & $0.48$ \\ 
\hline
\end{tabular}
\end{table}

\section{Limits on neutrino fluences from GW events involving neutron stars}

The limits on $\nu_{e,x}$ and $\bar{\nu}_{e,x}$ fluence were alternatively obtained for NSNS and NSBH GW~events. 
The intervals for the analysis $(E_{1,2,3}, E_{emax})$ are set to the same way as previously described. 
The values of the monoenergetic neutrino energy $E_\nu$ and the supernova neutrino mean energy $\langle E \rangle$ from 0.5 to 50 MeV in increments of 0.5 MeV, the expected spectra of recoil electrons and the number of detected events in the interval $(E_{1,2,3}, E_{emax})$ were used in the calculations. 

\begin{figure}
\includegraphics[width=9cm, height=10cm]{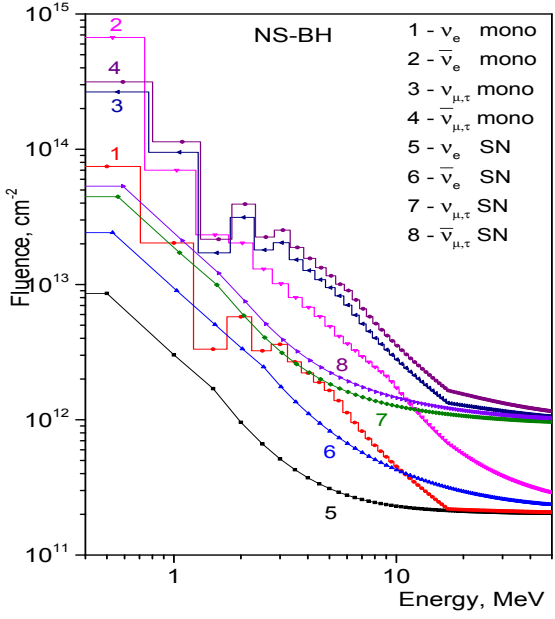}
	\caption{The merger of neutron star and black hole - GW190426 and GW191219 events. The upper limits on the fluences of monoenergetic neutrinos (Energy = $E_\nu$): 1-$\nu_e$, 2-$\bar{\nu}_e$, 3-$\nu_{\mu,\tau}$,  4-$\bar{\nu}_{\mu,\tau}$ and neutrinos with supernova spectra (Energy=$\langle E \rangle$): 5-$\nu_e$, 6-$\bar{\nu}_e$, 7-$\nu_{\mu,\tau}$,  8-$\bar{\nu}_{\mu,\tau}$ (90\% C.L.).}\label{fig:6}
\end{figure}

The relation (\ref{eq:6}) was converted into the obtained fluence limits for all neutrino flavors given in Figure~\ref{fig:5} for NSNS GW~170817 event  and in Figure~\ref{fig:6} for NSBH GW~190426 and GW~191219 events, both for mono-energetic neutrinos and for supernova neutrinos. 
We only analyzed the first NSNS GW~170817 event separately, since the second NSNS GW~190425 event occurred at a distance 4~times further, and the exactly GW~170817 event was analyzed by almost all of the above-mentioned neutrino detectors.

At neutrino energies above $17$~MeV, the limits become almost constant since the $(\nu,e)$-scattering cross section is proportional to $E_\nu$ and the spectrum of recoil electrons weakly depends on the electron energy.

\begin{table}[ht]
\caption{Borexino $90$\%~C.L. upper limits for the fluences of all neutrino flavours from NSNS GW~170817 event, obtained through the study of $(\nu,e)$ elastic scattering of monoenergetic neutrinos and IBD reaction. $E_{\nu}$ is given in MeV units, ${\Phi}_{\nu_{e,x},\bar{\nu}_{e,x}}$ -- in $10^{12}~\rm{cm^{-2}}$ units. }
\label{tab:4}
\begin{tabular}{|c|c|c|c|c|c|}
\hline
$E_{\nu}$ & ${\Phi}_{\nu_e}$ & ${\Phi}_{\bar{\nu}_{e}}$ & ${\Phi}_{\nu_{\mu,\tau}}$ & ${\Phi}_{\bar{\nu}_{\mu,\tau}}$ & ${\Phi}_{\bar{\nu}_{e}}$IBD\\
\hline
2 & $8.4$ & $23$  & $44$ & $53$ & $3.8 $ \\
6 & $1.9$ & $4.7$ & $11$ & $13$ & $0.062 $ \\
10 & $0.95$ & $2.6$ & $5.9$ & $7.4$ & $0.019 $ \\
14   & $0.59$ & $1.8$ & $3.6$ & $4.5$ & $0.0090 $ \\
18   & $0.46$ & $1.3$ & $2.8$ & $3.4$ &$ - $ \\
30   & $0.44$ & $0.81$ &$2.4$ & $2.8$ & $ - $\\
50   & $0.44$ & $0.61$ & $2.2$ & $2.4$ & $ - $ \\
\hline
\end{tabular}
\end{table}

The results for the NSNS case are also shown in Table~\ref{tab:4} for mono-energetic neutrinos. 
In case of the supernova neutrino spectrum, the fluence constraints are slightly stronger than the monoenergetic neutrino limit at $\langle E \rangle = E_\nu$ (Table~\ref{tab:5}). 
The wide expected neutrino spectrum allows to set limits at  energies $\langle E \rangle \geq 16.8$~MeV.

\begin{table}[ht]
\caption{Borexino $90$\%~C.L. upper limits for the fluences of all neutrino flavours from NSNS GW~170817 event, obtained through the study of $(\nu,e)$ elastic scattering of neutrino with supernova spectrum and IBD reaction. $\langle E \rangle$ is given in MeV units, ${\Phi}_{\nu_{e,x},\bar{\nu}_{e,x}}$ -- in $10^{12}~\rm{cm^{-2}}$ units. }
\label{tab:5}
\begin{tabular}{|c|c|c|c|c|c|}
\hline
$\langle E \rangle$ & ${\Phi}_{\nu_e}$ & ${\Phi}_{\bar{\nu}_{e}}$ & ${\Phi}_{\nu_{\mu,\tau}}$ & ${\Phi}_{\bar{\nu}_{\mu,\tau}}$ & ${\Phi}_{\bar{\nu}_{e}}$IBD\\
\hline
2 & $1.8$ & $4.4$  & $10$ & $12$ & $1.7 $ \\
6 & $0.58$ & $1.4$ & $3.4$ & $4.0$ & $0.12 $ \\
10 & $0.48$ & $0.89$ & $2.7$ & $3.0$ & $0.046 $ \\
14   & $0.46$ & $0.73$ & $2.4$ & $2.7$ & $0.040 $ \\
18   & $0.45$ & $0.65$ & $2.3$ & $2.5$ &$0.037$ \\
30   & $0.43$ & $0.55$ &$2.1$ & $2.3$ & $ 0.069 $\\
50   & $0.43$ & $0.50$ & $2.0$ & $2.2$ & $ 0.21 $ \\
\hline
\end{tabular}
\end{table}

\begin{figure}
	\includegraphics[width=9cm, height=10cm]{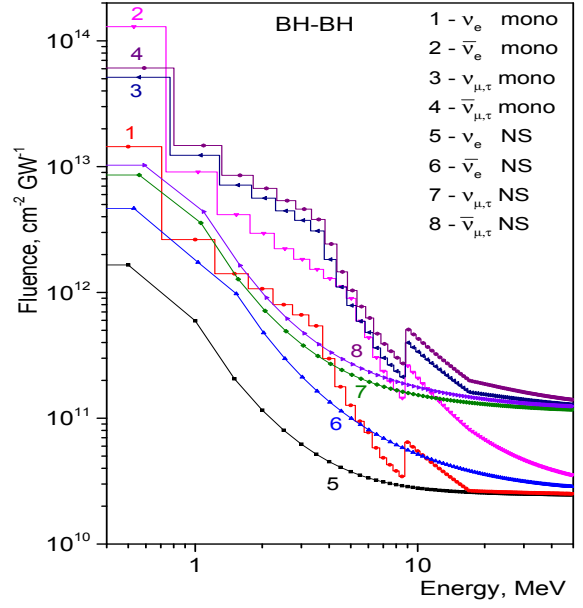}
	\caption{The upper limits on the neutrino fluences from 26 most intensive HBHB merges. The mono-energetic neutrinos (Energy=$E_\nu$): 1-$\nu_e$, 2-$\bar{\nu}_e$, 3-$\nu_{\mu,\tau}$,  4-$\bar{\nu}_{\mu,\tau}$. The neutrinos with supernova spectra (Energy=$\langle E \rangle$): 5-$\nu_e$, 6-$\bar{\nu}_e$, 7-$\nu_{\mu,\tau}$,  8-$\bar{\nu}_{\mu,\tau}$. (all for 90\% C.L.).}\label{fig:7}
\end{figure}

\section {Limits on the neutrino fluences from the selected most intensive BHBH GW~events.}
Since for almost all GW~events the distance to the event $R_i$ and the radiation mass $M^{rad}_i$ are known (with some accuracy), in contrast to gamma-ray bursts or fast radio bursts cases~\cite{Ago2017a,App2022}, it is possible to obtain the limit on the neutrino fluence for some hypothetical standard event at a fixed distance and with a known radiative mass. 
This allows to compare the obtained limits on the neutrino fluence for a different set of GW~events.
The neutrino fluence from i-th GW~event assuming isotropic emission is proportional:
\begin{equation}\label{eq:10}
    \Phi_i = \frac{\epsilon M^{rad}_i c^2} {4\pi R_i^2 \langle E_i \rangle},
\end{equation}
where $\epsilon$ is a fraction of neutrino radiation and $\langle E_i \rangle$ is the average neutrino energy. 
The obtained limits on the neutrino fluences can be converted to the limit on the value of $\epsilon$. 

According to the GWTC-3 database for 70 BHBH  GW~events when the Borexino detector was in data taking mode, the distances to events $R_i$ lie in the range  (0.3 - 8.3 Gpc) and an average distance  is $\langle R_i \rangle=~$2.1 Gpc. 
The corresponding radiative masses $M^{rad}_i$ are inside (0.3 - 9.4) $M_\odot$ interval with average mass $\langle M^{rad}_i \rangle = 2.6  M_\odot$. 
Therefore, as a reference event, we define the "standard" event with $M^{rad}$ equals one solar mass $M_\odot$  at a distance of one Gpc .
The upper limits on the neutrino fluence obtained for the single GW~event $\Phi_i$ can be converted to the limit on the fluence from "standard" GW~event as $\Phi_{std}=w_i\Phi_i$ where the weight $w_i = (1~\rm{Gpc}/R_i)^2 (M^{rad}_i/M_\odot)$.

Among 70 BHBH GW~events, we selected 26 events, for which weight $w_i$ is greater than 1.
For these 26 events, the average distance is 1.0~Gpc,  the average radiative mass is 2.3~$M_\odot$, and the factor characterizing the neutrino flux  $\langle w_{26} \rangle$=3.1.
Correspondingly, for the remaining 44 events, the average distance is 2.8 Gpc, the radiative mass is 2.9~$M_\odot$, and the factor $\langle w_{43} \rangle$ = 0.44.
These 26 most intense events were used to search for an additional contribution to the $\pm~1000$~s temporal interval from the neutrino interactions, the spectrum was represented as mono-energetic neutrinos and neutrinos with a supernova spectrum. 
Since no statistically significant excesses were found, the resulting upper limits on the neutrino fluences for all flavors are shown in the Figure~\ref{fig:7}.
Although the upper bounds on the fluences for mono-energetic neutrinos in Figure~\ref{fig:7} (line 1-4) turn out to be weaker than in Figure~\ref{fig:4}, they are obtained for GW~events with the expected most intense neutrino fluxes.

\section {Limits on the ${\bar{\nu}_e}$-fluence from the IBD reaction}
\label{section:IBD}
As already mentioned, electron antineutrinos can be also detected in the Borexino detector via inverse $\beta$-decay (IBD) reaction on protons with energy threshold of $E_{\bar{\nu}_e}=1.8$~MeV.
The cross section of this process is much higher than the one for $({\bar{\nu}_e},e)$ elastic scattering. 
Additionally, the IBD offers a unique signature given by temporal and spatial coincidence of two correlated events associated with detection of a positron and a neutron. 
The prompt positron event with visible energy of $E_{\bar{\nu}_e} - 0.784$~MeV accompanied by $\gamma$-rays from neutron capture mostly on protons or carbon nuclei with a small probability.
As a result, the rate of the events selected as IBD candidates is much lower with respect to the rate of single electron-like events.

The procedure of IBD events selection and the energy spectrum of prompt positron events are described in detail in~\cite{Bel2013g,Ago2015,Ago2020b,Bel2011c,Ago2021}.
In the IBD analysis, we used the same 16.8~MeV upper boundary of the visible energy range as in the case of the $(\nu,e)$-scattering analysis.
No IBD events were observed in the $\pm 1000$~s interval around the selected GW~events and the expected background was almost zero~\cite{Ago2020b,Ago2021} that allowed us to use the conservative value of $N_{90}(E_\nu,n_\mathrm{obs},n_\mathrm{bkg})$ = 2.44 in the analysis~\cite{Fel1998}. Since the cross section of IBD reaction is about two orders of magnitude larger than $(\nu,e)$-scattering cross sections at given neutrino energies, and the background level is smaller, the most stringent upper limits have been obtained for the fluence of electron antineutrinos.

\begin{figure}
	\includegraphics[width=9cm, height=10cm]{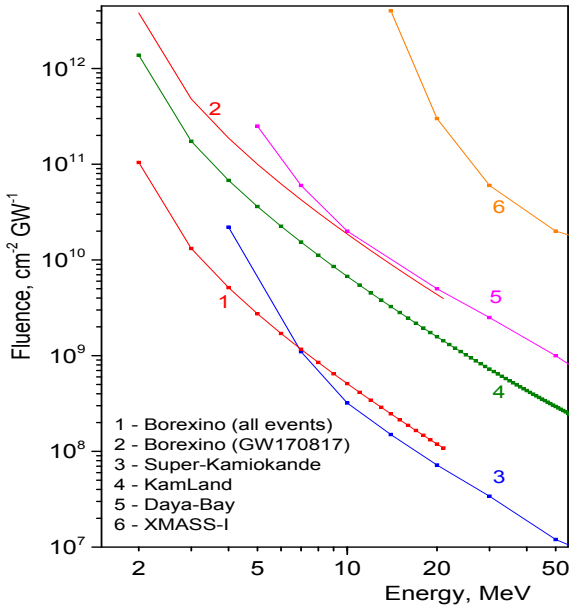}
	\caption{Upper limits on the fluence of mono-energetic electron antineutrinos $\bar{\nu}_e$ obtained using IBD reaction. 1-Borexino (all 74 GW~events, reduced per single GW~event), 2-Borexino (GW~170817 NSNS event), 3-Super-Kamiokande coll.~\cite{Abe2018} (GW~170817), 4-KamLand coll.~\cite{Abe2021A} (GW~170817), 5- DayaBay coll.~\cite{An2021}, 6 - XMASS-I coll.~\cite{Abe2021} (all for  90\% C.L.).}\label{fig:8}
\end{figure}

The upper limit on the mono-energetic electron antineutrinos (${\bar{\nu}_e}$) fluence using the IBD reaction is calculated from the relation (\ref{eq:6}) but with replacing of the $N_e$ with the number of protons $N_p$ and considering the cross section of IBD reaction~\cite{Str2003}. 
The resulting limit on ${\bar{\nu}_e}$ fluence for 74 GW~events reduced per one GW~event is shown in Figure~\ref{fig:8} (line 1) and Tables~\ref{tab:3},~\ref{tab:4} and~\ref{tab:5}. 
For comparison, the limit on $\bar{\nu}_e$ fluence for one GW~event (e.g. GW170817 NSNS merge) is presented by line 2 in Figure~\ref{fig:8}.

Upper limits on the fluence can be converted into upper limits on the total energy radiated in form of neutrinos $\epsilon M^{rad}_i$  for a BHBH, NSNS and NSBH mergers using relation (\ref{eq:10}). 
We consider only the energy radiated by $10$~MeV electron neutrinos $\nu_e$ and antineutrinos $\bar{\nu}_e$ under  assumption of isotropic angular distribution of emitted neutrinos.
Upper limits on the fluence of $\nu_e$ and $\bar{\nu}_e$ from the closest NSNS GW~170817 event obtained for ($\nu_e,e$)-scattering and from IBD reaction (\ref{tab:4}) lead to restrictions on $\epsilon M^{rad} c^2 \leq 2.9 \times 10^{60}$~erg and $\epsilon M^{rad} c^2 \leq 5.7 \times 10^{58}$~erg, accordingly.
These values can be compared with the energy of solar mass $M_\odot c^2 = 1.8\times 10^{54}$~erg.
Accordingly, the restrictions on fraction of neutrino radiation become unnatural and exceed unity ($\epsilon_{\nu_e} \leq 1.0\times 10^{9}$ and $\epsilon_{\bar{\nu}_e} \leq 2.0\times 10^{7}$).
For comparison, we note that if GW170817 event occurred at the same distance as the supernova collapse SN1987A at $51$~kpc from Earth, the sensitivity to the energy released in form of $10$~MeV neutrinos would be $\epsilon M^{rad} c^2 \leq 8.9 \times 10^{52}$~erg ($0.05 M\odot$) and $\epsilon_{\bar{\nu}_e} \leq 1.2$.

In the case of 26 most intensive BHBH GW~events, the resulting fluence is:
\begin{equation}\label{eq:11}
 \Phi_{tot} =\sum\limits_{i=1}^{26} \frac{\epsilon M^{rad}_i c^2} {4\pi R_i^2 <E_i>}.
\end{equation}
where the upper limit on the total fluence $\Phi_{tot} = 1.45\times 10^{12} \rm{cm^{-2}}$ (Figure~\ref{fig:7}) and values $M_{rad_i}$ and $R_i$ can be taken from the GWTC-3 database.
Since for 26 GW~events the average mass $\langle M^{rad} \rangle = 11 M_\odot$ as well as the average squared distance $\langle R^2 \rangle = 0.5~\rm{Gpc^2}$ are about 300 times larger than for the GW~170817 event, the upper limits of the value of $\epsilon_{\nu_e,\bar{\nu}_e}$ are of the same order as for obtained with GW~170917 event.


This suggests that successful detection of low-energy neutrinos should be possible only in case of anisotropic angular distribution of neutrino emission. 
The limits on energy radiated into neutrinos of other flavors can be easily calculated from Tables~\ref{tab:3},~\ref{tab:4} and~\ref{tab:5}.

\section{Conclusion}
We have searched for signals of neutrino-electron scattering with visible energies above $250$~keV within a time window of $\pm1000$~s centered at the detection moment of a particular GW~event. 
Two types of incoming neutrino spectra were considered: a mono-energetic line and supernova neutrino spectrum given by modified Fermi-Dirac distribution for different effective neutrino temperatures.
We searched for coincident neutrino-electron elastic scattering of $\nu_{e,\mu,\tau}$ and $\bar{\nu}_{e,\mu,\tau}$ and IBD of $\bar{\nu}_e$ in the Borexino detector with the $74$~GW~events associated with the O1, O2 and O3 observing runs of the LIGO/VIRGO detectors.
We looked for an excess in the number of Borexino events produced by neutrino-electron elastic scattering and the inverse beta-decay on protons correlated to $74$~GW~events from the GWTC-3 database.
We found no statistically significant increase in the number of events, with visible energies above $0.25$~MeV within time windows of $\pm1000$~s centered at the moment of GW arrival for all three options for merging of black holes and neutron stars. 
As a result, new limits on the fluence of monochromatic and supernova neutrinos of all flavors were set for neutrino energies in range of $0.5 - 5$~MeV. 
Also, the inverse beta-decay reaction was considered in order to set a new limit on the fluence of electron antineutrinos related with the GW~events.  

\begin{acknowledgements}
The Borexino program is possible by funding from INFN (Italy), NSF  (USA), DFG and HGF (Germany), RSF (Grant 21-12-00063) (Russia), and NCN (Grant No. UMO-2017/26/M/ST2/00915) (Poland). 
We acknowledge the generous hospitality and support of the Laboratori Nazionale del Gran Sasso (Italy).
\end{acknowledgements}



\end{document}